Thermal processes generated in QGP by yoctosecond ($10^{-24}$ s) laser pulses

J. Marciak-Kozłowska [1]

M. Kozłowski [2,*]

[1] Institute of Electron Technology, Warsaw

[2] Physics Department, Warsaw University, Warsaw

[*] corresponding author, e-mail: miroslawkozlowski@aster.pl

**Abstract**

In this paper the thermal processes generated by yoctosecond ($10^{-24}$ s) laser pulses in QGP are investigated. Considering that the relaxation time in QGP is of the order of 1 ys it is shown that in QGP the yoctosecond laser pulses can generate the thermal waves with velocity $v = c$ (0.3 fm/ys).

**Key words**: QGP, thermal waves, yoctosecond pulses

1. **Introduction**

The QGP (Quark Gluon Plasma) is formed in collisions of hadrons.

In this paper we develop the model for the thermal energy transport in QGP generated by yoctosecond laser pulses (1ys = $10^{-24}$ s = 0.3 fm/$c$). As was shown by A. Ipp et al., Phys. Rev. [1] the laser with yoctosecond photon pulses can be arranged in RHIC and LHC hadron colliders. In QGP the thermal relaxation time is of the order of ys. It means that for ys photons the duration time of the laser pulse and relaxation time is the same order. For those thermal processes the Fourier approximation – parabolic thermal diffusion is not valid. In this paper we compare both models – hyperbolic and parabolic and conclude that in the case of the hyperbolic model the new mode of QGP excitation – thermal wave can be generated.

## 2. Heaviside and Klein – Gordon thermal equation

First of all let us consider, for the moment, the parabolic heat transport equation [2]

$$\frac{\partial T}{\partial t} = \frac{\hbar}{m}\nabla^2 T. \tag{2.1}$$

When the real time $t \to it/2$, $T \to \Psi$, is the quantum wave function, Eq. (2.1) has the form of a free Schrödinger equation

$$i\hbar\frac{\partial \Psi}{\partial t} = -\frac{\hbar^2}{2m}\nabla^2\Psi. \tag{2.2}$$

The complete Schrödinger equation has the form

$$i\hbar\frac{\partial \Psi}{\partial t} = -\frac{\hbar^2}{2m}\nabla^2\Psi + V\Psi, \tag{2.3}$$

where $V$ denotes the potential energy. When we go back to real time $t \to 2it$, $\Psi \to T$, the parabolic quantum heat transport is obtained

$$\frac{\partial T}{\partial t} = \frac{\hbar}{m}\nabla^2 T - \frac{2V}{\hbar}T. \tag{2.4}$$

Equation (2.4) describes the quantum heat transport for $\Delta t > \tau$. For heat transport initiated by yoctosecond laser pulses, $\Delta t < \tau$, one obtains the generalized quantum hyperbolic heat transport equation with memory term added

$$\tau\frac{\partial^2 T}{\partial t^2} + \frac{\partial T}{\partial t} = \frac{\hbar}{m}\nabla^2 T - \frac{2V}{\hbar}T. \tag{2.5}$$

Considering that $\tau = \dfrac{\hbar}{m\upsilon^2}$, Eq. (2.5) can be written as follows:

$$\frac{1}{\upsilon^2}\frac{\partial^2 T}{\partial t^2} + \frac{m}{\hbar}\frac{\partial T}{\partial t} + \frac{2Vm}{\hbar^2}T = \nabla^2 T. \tag{2.6}$$

Equation (2.6) describes the heat flow when besides the temperature gradient, the potential energy $V$ operates.

In the following, we consider one-dimensional heat transfer phenomena, i.e

$$\frac{1}{\upsilon^2}\frac{\partial^2 T}{\partial t^2} + \frac{m}{\hbar}\frac{\partial T}{\partial t} + \frac{2Vm}{\hbar^2}T = \frac{\partial^2 T}{\partial x^2}. \tag{2.7}$$

For quantum heat transfer equation (2.7) we seek the solution in the form

$$T(x,t) = e^{\frac{1}{2}}\, u(x,t) \tag{2.8}$$

After substitution (2.8) into Eq. (2.7), one obtains

$$\frac{1}{^2}\frac{\partial^2 u}{\partial t^2} - \frac{\partial^2 u}{\partial x^2} + qu(x,t) = 0, \tag{2.9}$$

where

$$q = \frac{2Vm}{\hbar^2} - \left(\frac{m\upsilon}{2\hbar}\right)^2. \tag{2.10}$$

In the following, we will consider a constant potential energy $V = V_0$. The general solution of Eq. (2.9) for the Cauchy boundary conditions [2],

$$u(x,0) = f(x), \qquad \left[\frac{\partial u(x,t)}{\partial t}\right]_{t=0} = F(x), \tag{2.11}$$

has the form [2]

$$u(x,t) = \frac{f(x-\upsilon t) + f(x+\upsilon t)}{2} + \frac{1}{2\upsilon}\int_{x-\upsilon t}^{x+\upsilon t}\Phi(x,y,z)dz, \tag{2.12}$$

where

$$\Phi(x,t,z) = \frac{1}{\upsilon}F(z)J_0\left(\frac{b}{\upsilon}\sqrt{(z-x)^2 - \upsilon^2 t^2}\right) + btf(z)\frac{J_0'\left(\frac{b}{\upsilon}\sqrt{(z-x)^2 - \upsilon^2 t^2}\right)}{\sqrt{(z-x)^2 - \upsilon^2 t^2}},$$
$$b = \left(\frac{m\upsilon^2}{2\hbar}\right)^2 - \frac{2Vm}{\hbar^2}\upsilon^2 \tag{2.13}$$

and $J_0(z)$ denotes the Bessel function of the first kind. The function $u(x,t)$ describes the propagation of the distorted thermal quantum waves with characteristic lines $x = \pm \upsilon t$.

We can define the distortionless thermal wave as the wave that preserves the shape in the field of the potential energy $V_0$. The condition for conserving the shape can be formulated as

$$q = \frac{2Vm}{\hbar^2} - \left(\frac{m\upsilon}{2\hbar}\right)^2 = 0. \tag{2.14}$$

When Eq. (2.14) holds, Eq. (2.10) has the form

$$\frac{\partial^2 u(x,t)}{\partial t^2} = \upsilon^2 \frac{\partial^2 u}{\partial x^2}. \tag{2.15}$$

Equation (2.15) is the quantum thermal wave equation with the solution (for Cauchy boundary conditions (2.11))

$$u(x,t) = \frac{f(x-\upsilon t) + f(x+\upsilon t)}{2} + \frac{1}{2\upsilon}\int_{x-\upsilon t}^{x+\upsilon t} F(z)dz. \tag{2.16}$$

It occurs that structure of equation (2.9) depends on the sign of the parameter $q$. For quantum heat transport, e.g. in nuclear matter, parameter $q$ is the function of potential barrier height $V_0$.

In monograph [1] the velocity of thermal waves in nuclear (protons and neutrons) and in QGP was calculated

$$\upsilon_i = \alpha_i c, \qquad i = 1, 2 \tag{2.17}$$

where $c$ is the light velocity and $\alpha_1 = 0.16$ for nuclear matter and $\alpha_2 = 1$ for QGP.

Respectively we obtain for thermal relaxation time

$$\tau_i = \frac{\hbar}{m_i \upsilon_i^2}, \qquad i = 1, 2 \tag{2.18}$$

and

$$\tau_1 = \frac{\hbar}{m_h (\alpha_1 c)^2}, \qquad \tau_2 = \frac{\hbar}{m_q c^2}.$$

For Cauchy initial condition [2]

$$u(x,0) = f(x), \qquad \frac{\partial u(x,0)}{\partial t} = g(x)$$

the solution of Eq. (2.10) for $q<0$ has the form

$$u(x,t) = \frac{f(x-\upsilon t) + f(x+\upsilon t)}{2}$$
$$+ \frac{1}{2\upsilon} \int_{x-\upsilon t}^{x+\upsilon t} g(\varsigma) I_0\left[\sqrt{-q\left(\upsilon^2 t^2 - (x-\varsigma)^2\right)}\right] d\varsigma \qquad (2.19)$$
$$+ \frac{\left(\upsilon\sqrt{-q}\right)t}{2} \int_{x-\upsilon t}^{x+\upsilon t} f(\varsigma) \frac{I_1\left[\sqrt{-q\left(\upsilon^2 t^2 - (x-\varsigma)^2\right)}\right]}{\sqrt{\upsilon^2 t^2 - (x-\varsigma)^2}} d\varsigma$$

and the equation (2.10) is the modified Heaviside equation. When $q>0$ equation (2.10) is the modified Klein – Gordon thermal equation. The solution of M K-G equation has the same structure as the formula (2.19) but with $q \to -q$ and $I_0(z), I_1(z) \to J_0(z), J_1(z)$ [2].

### 3. The model thermal quantum equation

For free thermal energy transport equation (2.7) can be written as:

$$\frac{1}{\upsilon^2} \frac{\partial^2 T}{\partial t^2} + \frac{m}{\hbar} \frac{\partial T}{\partial t} = \frac{\partial^2 T}{\partial x^2}. \qquad (3.1)$$

In the subsequent we are concerned with the solution to (3.1) for a nearly delta function temperature generated in nuclear matter. The pulse transfered has the shape:

$$\begin{aligned} \Delta T_0 & \quad \text{for } 0 < x < \Delta l, \\ \Delta T_0 = 0 & \quad \text{for } x > \Delta l. \end{aligned} \qquad (3.2)$$

With $t = 0$ temperature profile the solution of equation (3.1) is [2]:

$$T(l,t) = \tfrac{1}{2}\Delta T_0 e^{-t/2\tau} \Theta(t-t_0)\Theta(t_0 + \Delta t - t)$$
$$+ \tfrac{\Delta t}{4\tau} \Delta T_0 e^{-t/2\tau} \left\{ I_0(z) + \frac{t}{2\tau}\frac{1}{z} I_1(z) \right\} \Theta(t-t_0), \qquad (3.3)$$

where $z = \left(t^2 - t_0^2\right)^{1/2}/2$ and $t_0 = l/\upsilon_S$. The solution to eq.(3.3) when there are reflecting boundaries is the superposition of the temperature at $l$ from the original temperature and from image heat sources at $\pm 2nl$:

$$T(l,t) = \sum_{i=0}^{\infty} \left[ \begin{array}{l} \Delta T_0 e^{-t/2\tau}\Theta(t-t_i)\Theta(t_i + \Delta t - t) \\ +\Delta T_0 \dfrac{\Delta t}{2\tau} e^{-t/2\tau}\left\{ I_0(z_i) + \dfrac{t}{2\tau}\dfrac{1}{z_i} I_1(z_i) \right\} \Theta(t-t_i) \end{array} \right], \qquad (3.4)$$

In Eqs (3.3) and (3.4) $\Theta(y)$ is the Heaviside step function.

### 4. Yoctosecond photon pulses from quark – gluon plasma

Among the shortest possible time scales that are available experimentally are those obtained through high energy collisions. Particularly interesting in this context are heavy ion collisions that can produce a quark – gluon plasma (QGP). Heavy ion collision at the Relativistic Heavy Ion Collider (RHIC) and LHC at CERN produce this new state of matter up to the size of nucleous (~ 15 fm) for a duration of a few tens of yoctoseconds (1ys = $10^{-24}$ s = 0.3 fm/$c$) [3]. In such a collision the QGP is produced initially in a very anisotropic state and reaches a hydrodynamic evolution through internal interactions only after a few relaxation time . The observed particle spectra turned out to agree well with ideal hydrodynamical model predictions which led to the assumption that relaxation time is of the order of 1 ys.

In this paragraph we calculate the temperature field $T(x,t)$ for hadron collisions following the model presented in paragraph 2. In monograph [2] the relaxation time for QGP was calculated:

$$\tau = \frac{\hbar}{m_q \left(\alpha_q c\right)^2} \approx \sim 10^{-24} \text{s} = 1\text{ys}, \quad (4.1)$$

where $\alpha_q = 1$, $m_q = 417$ MeV and $\hbar$ is the Planck constant.

In Fig. 1-3 the results of calculations are presented. For all figures the duration of laser pulse is 1 ys = 0.3 fm/$c$. In Fig. 1 a, b the relaxation time is 5 fm/$c$ and $v = c$. Fig. 1 a presents the result of calculation with Klein – Gordon thermal relaxation time. The temperature field has the structure of thermal wave with velocity $c$. the Fourier model (parabolic equation) is presented in Fig. 1 b. In Fig. 2, 3 the calculations ere presented for relaxation time 1 fm/$c$ and 0.5 fm/$c$ respectively.

### 5. Conclusions

In this paper the interaction of yoctosecond laser pulses with QGP is investigated. It is shown that for relaxation time in the range 0.5 – 5 fm/*c* (1 –10 ys) in QGP the thermal waves with velocities can be created.

**Figure captions**:

Fig. 1 a. Temperature field *T*(*x*,*t*). Parameters: relaxation time = 5 fm/*c*, *v* = *c*. Hyperbolic model calculation.

Fig. 1 b. Temperature field *T*(*x*,*t*). Parameters: relaxation time = 5 fm/*c*, *v* = *c*. Parabolic model calculation.

Fig. 2 a,b The same as in Fig. 1 a,b but = 1 fm/*c*.

Fig. 3 a,b The same as in Fig. 1 a,b but = 0.5 fm/*c* 1 ys.

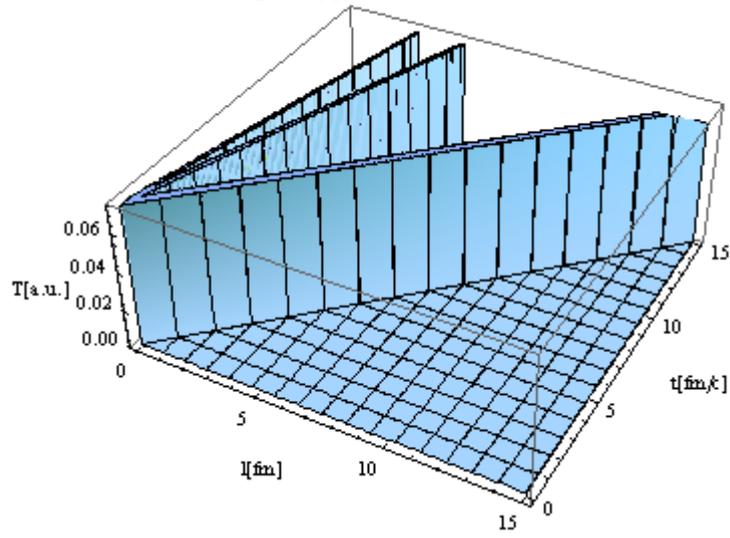

Fig.1a Hyperbolic model tau=5 fm/c

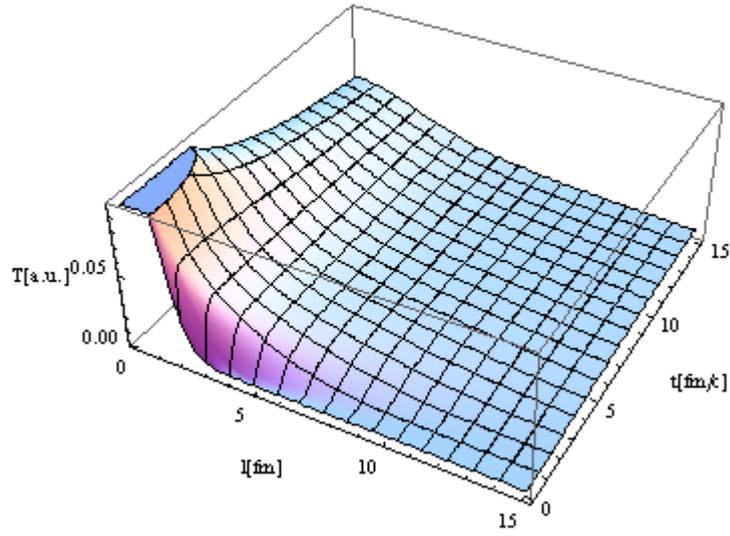

Fig.1 b Parabolic Model tau=5 fm/c

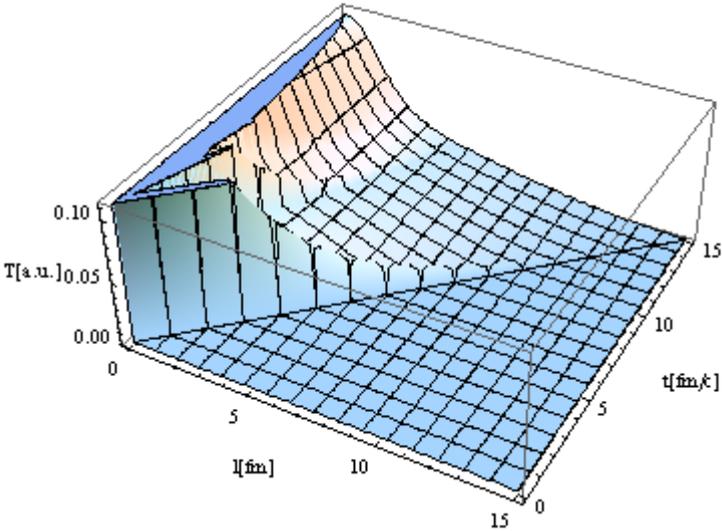

Fig.2 a Hyperbolic model tau=1 fm/c

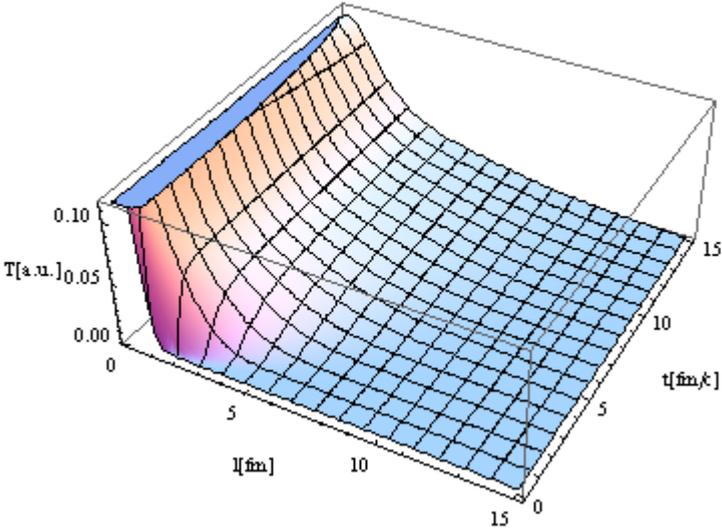

Fig.2 b Parabolic Model tau=1 fm/c

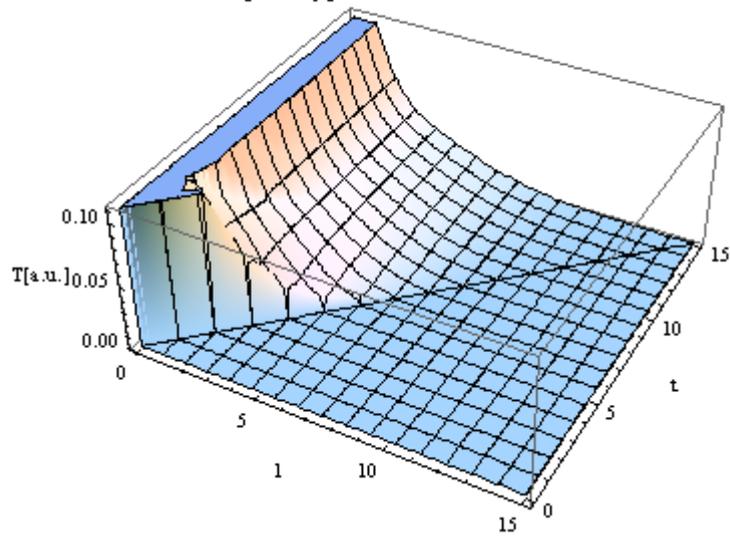

Fig.3 a Hyperbolic model tau=0.5 fm/c

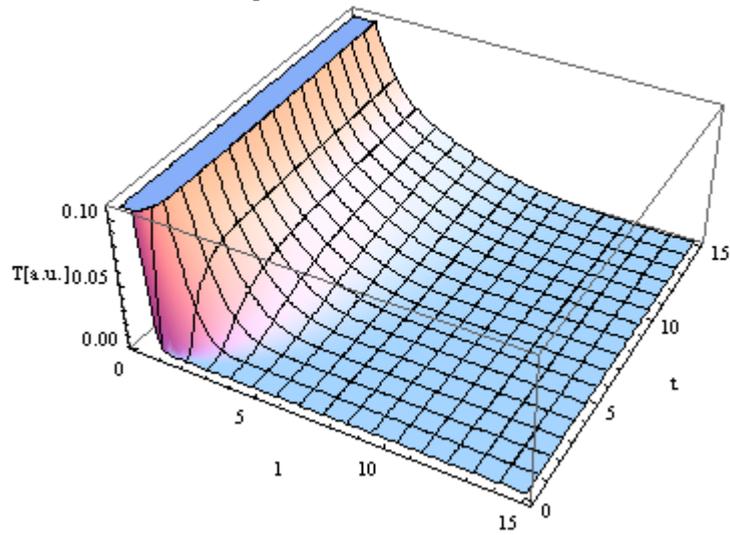

Fig.3 b Parabolic Model tau=0.5 fm/c